\documentclass[sn-mathphys-num]{sn-jnl}

\usepackage{graphicx}%
\usepackage{multirow}%
\usepackage{amsmath,amssymb,amsfonts}%
\usepackage{amsthm}%
\usepackage{mathrsfs}%
\usepackage[title]{appendix}%
\usepackage{xcolor}%
\usepackage{textcomp}%
\usepackage{manyfoot}%
\usepackage{booktabs}%
\usepackage{algorithm}%
\usepackage{algorithmicx}%
\usepackage{algpseudocode}%
\usepackage{listings}%

\theoremstyle{thmstyleone}%
\theoremstyle{thmstyletwo}%

\theoremstyle{thmstylethree}%

\begin{document}

\title[Loewner Evolution for Critical Invasion Percolation Tree]{Loewner Evolution for Critical Invasion Percolation Tree}


\author*[1]{\fnm{Leidy M. L.} \sur{Abril}}\email{leidy@fisica.ufc.br}

\author[1]{\fnm{André A.} \sur{Moreira}}\email{auto@fisica.ufc.br}

\author[1]{\fnm{José S.} \sur{Andrade Jr.}}\email{soares@fisica.ufc.br}

\author[1,2]{\fnm{Hans J.} \sur{Herrmann}}\email{hans@fisica.ufc.br}

\affil*[1]{\orgdiv{Departamento de Física}, \orgname{Universidade Federal do Ceará}, \orgaddress{\street{Campus do Pici}, \city{Fortaleza}, \postcode{60451-970}, \state{Ceará}, \country{Brazil}}}

\affil[2]{\orgdiv{PMMH}, \orgname{ESPCI}, \orgaddress{\street{7 quai St. Bernard}, \city{Paris}, \postcode{75005}, \country{France}}}

\abstract{Extending the Schramm--Loewner Evolution (SLE) to model branching structures while preserving conformal invariance and other stochastic properties remains a formidable research challenge. Unlike simple paths, branching structures, or trees, must be associated with discontinuous driving functions. 
Moreover, the driving function of a particular tree is not unique and depends on the order in which the branches are explored during the SLE process.
This study investigates trees formed by nontrapping invasion percolation (NTIP) within the SLE framework. Three strategies for exploring a tree are employed: the invasion percolation process itself, Depth--First Search (DFS), and Breadth--First Search (BFS).
We analyze the distributions of displacements of the Loewner driving functions and compute their spectral densities. Additionally, we investigate the inverse problem of deriving new traces from the driving functions, achieving a reasonably accurate reconstruction of the tree-like structures using the BFS and NTIP methods. Our results suggest the lack of conformal invariance in the exploration paths of the trees, as evidenced by the non-Brownian nature of the driving functions for the BFS and NTIP methods, and the inconsistency of the diffusion constants for the DFS method.}




\maketitle

\section{Introduction}\label{sec:introduction}

Invasion percolation models are important for describing fluid transport in porous media, with numerous applications, including oil and gas extraction from reservoirs~\cite{etienne2022}, groundwater hydrology~\cite{berkowitz1993percolation}, and contaminant transport in soils~\cite{tsakiroglou2008dual}, among others~\cite{LOPEZ200376,araujo2005,SABERI20151}.  
Étienne Guyon made significant contributions to this field, particularly in understanding the behavior of disordered systems and the role of connectivity in fluid flow through porous media~\cite{Wilke1985,guyon2020built,guyon2015physical}. 
He also explored how rock rupture emerges as a hidden percolation process~\cite{roux1988rupture}, investigated the local transport properties of fractured media~\cite{charlaix1987permeability} and dynamics of invasion percolation in various models~\cite{Roux_1989}. His insights into percolation theory have significantly advanced our understanding of the complex nature of materials and their transport properties.

To explore these phenomena, the porous material is modeled as a lattice of sites, where each site represents a pore. During the invasion process, the fluid penetrates the porous medium, forming clusters that can span the entire lattice.
At the percolation threshold, these clusters exhibit scale invariance, a property crucial for predicting flow patterns across different scales in such materials.

In two dimensions, a broader property--conformal invariance--has been observed at the percolation threshold in certain percolation models~\cite{nina2024,pose2014shortest,daryaei2012watersheds}. Conformal invariance, which preserves angles locally, contributes to the classification of critical models through a few universal parameters that define their universality classes~\cite{SABERI20151}. Determining whether conformal invariance exists in the invasion percolation model could offer valuable insights into the system's behavior, particularly with regard to fluid transport.
A powerful tool for addressing conformal invariance at the percolation threshold is the Schramm--Loewner Evolution (SLE) theory. 

The SLE theory successfully captures key scaling  properties of two--dimensional fractal curves in the continuum limit, with the diffusion coefficient $\kappa$ as the single parameter~\cite{schramm2000scaling}.
The core idea of SLE is that each curve is described as a local,  continuous growth process in the upper half--plane, which is then mapped into a real--valued `driving' function~\cite{bauer20062d,credidio2016stochastic,schramm2000scaling,gruzberg2004loewner,tizdast2022self}.

If the stochastic process that describes a curve
in the plane obeys conformal invariance and the domain Markov property, the driving function should correspond to a one--dimensional Brownian motion with diffusion coefficient $\kappa$. 
Examples of SLE curves, all of which are continuous paths, include self--avoiding walks~\cite{kennedy2002monte}, the boundary of the uniform spanning tree~\cite{bauer20062d}, and the boundary of critical site percolation~\cite{smirnov2001critical}. However, the theory does not account for systems with branching structures, such as branched polymers.

Motivated by exploring the connection between the SLE theory and tree-like structures, recent generalizations of the chordal Loewner evolution have been introduced~\cite{healey2023scalinglimitsbranchingloewner,Kemppainen2017,peltola2024loewnertracesdrivenlevy,oikonomou2008global}.
One approach to addressing branching structures without modifying the SLE theory is the local exploration process introduced by Sheffield~\cite{sheffield2009exploration}.
This exploration process is a depth--first search algorithm, ensuring that all sites are visited while avoiding loops.
In a similar spirit, our interest here lies in characterizing the driving functions of branching structures based on alternative exploration processes.  

One example of a branching structure is (loopless) Invasion Percolation (IP), which models fluids penetrating porous media and is important in oil exploration~\cite{sheppard1999invasion}.
The IP model, introduced by Wilkinson and Willemsen~\cite{wilkinson1983invasion} describes the displacement of a viscous fluid in a porous medium by another injected fluid~\cite{ebrahimi2010invasion}. In two dimensions, there are two variants of IP~\cite{sheppard1999invasion,schwarzer1999structural}: nontrapping and trapping IP. In the nontrapping variant (NTIP), the injected fluid can freely move into all areas of the viscous fluid, even if those areas are already encircled by the injected fluid. In contrast, in the trapping variant, the fluid cannot invade areas of the viscous fluid that are surrounded by the injected fluid. 

The NTIP model differs from standard percolation because it does not have a control parameter, instead, the cluster grows until it touches the edge of the lattice. NTIP is believed to belong to the same universality class as random percolation, as clusters in these models exhibit the same fractal dimension~\cite{sheppard1999invasion,knackstedt2000invasion,araujo2005}.
Nevertheless, the NTIP model has not yet been studied within the framework of SLE. 
For cross--validation, the parameter $\kappa$ can be computed using four independent numerical methods: the fractal dimension~\cite{beffara2008}, the winding angle~\cite{daryaei2012watersheds}, left--passage probability~\cite{schrammpercolation2001} and direct SLE. 
If all methods give the same value, the curve qualifies for belonging to the SLE family~\cite{pose2018schramm,nina2024}.

The paper is organized as follows: In Sec.~\ref{sec:Model}, we describe the algorithms used to generate and explore the trees:  NTIP,  DFS and,  BFS. Here, we also introduce the SLE theory, and explain the method for computing the driving function.
In Sec.~\ref{sec:results}, we study the statistics of the driving functions by analyzing their distributions and moments, computing the diffusion coefficient and determining the power spectral density. 
We also address the inverse problem of reconstructing the corresponding tree from a given driving function. The new trace is quantified by its fractal dimension. Finally, we focus on  the DFS method to determine the parameter $\kappa$ through three numerical tests and analyze its values. 
We conclude in Sec.~\ref{sec:conclusions}.

\section{Methods\label{sec:Model}} 

In this section, we introduce the nontrapping invasion percolation (NTIP) algorithm. Additionally, we describe the Depth--First Search~\cite{tarjan1972depth}  and Breadth--First Search~\cite{awerbuch1987new} algorithms, which are employed to define the order in which the branches of the tree are visited.  We also present the zipper algorithm, which maps the trees to driving functions~\cite{kennedy2007fast}.
Our goal is to investigate how different schemes of exploring the tree sculp the corresponding driving functions.

\subsection{Creating the tree \label{subsec:creating_tree}} 

\begin{figure}[b]
\includegraphics[width=\linewidth]{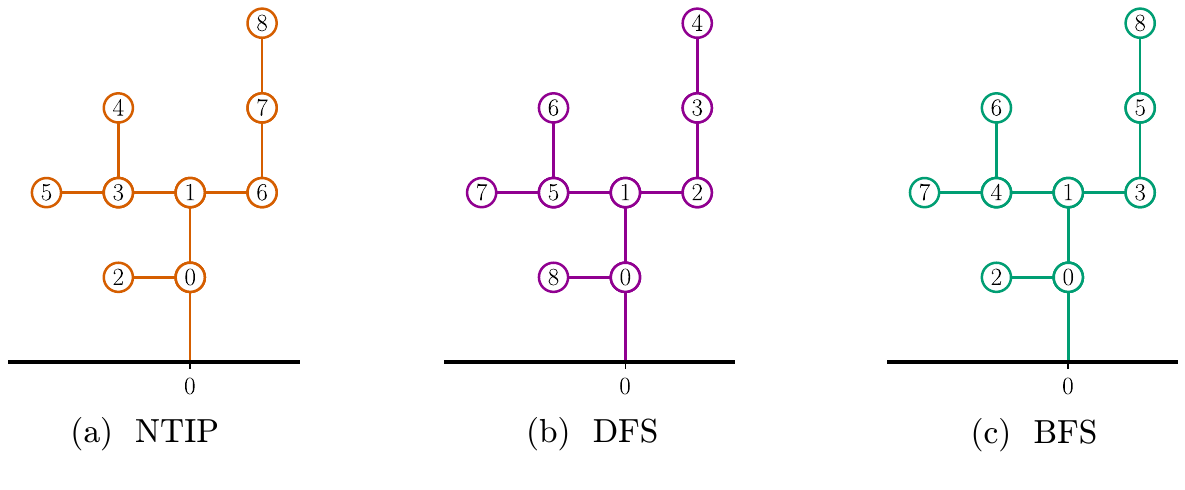}
\caption{ Exploring branching structures. We construct a tree on a rectangular lattice with $L=4$. 
In (a), the numbers indicate the order in which the sites were created during the NTIP process.  In (b), the same tree is shown, but the sites are visited in the order determined by the Depth--First Search (DFS) algorithm. DFS explores the branch furthest to the right before backtracking to visit other branches.  
In (c), the tree is explored using the Breadth--First Search (BFS) algorithm. BFS visits all sites at the same chemical distance from the origin before exploring the next chemical distance. 
}
\label{fig:trees}
\end{figure}

Here we give an outline of how the NTIP algorithm is defined: 
We create a tree on a square lattice in the upper half--plane $\mathbb{H}$, limiting the lattice to a size $2L-1  \times L$. The NTIP process starts at the origin also called root site, located at the center of the bottom boundary, labeled as node ‘0' in Fig.~\ref{fig:trees}.

Starting from the root, we assign uniformly distributed random numbers between 0 and 1 to its three nearest neighbors and include these sites in a list of accessible sites for the growing cluster. 
From this list, the site with the smallest assigned value is selected and marked as invaded. This site is then removed from the list of accessible sites, and a bond is formed between it and a neighbor that already belongs to the tree.
We move to the invaded site and repeat the process: assigning random values to all unvisited neighbors, adding them to the list of accessible neighbors and then again selecting the site with the smallest value. This iterative process continues until the growing cluster reaches either the sides or the upper boundary of the lattice. 

Given a specific tree generated by the NTIP algorithm, we must determine an order to go through its branches during the SLE process. In fact, the NTIP algorithm itself already provides an invasion order that can be used to obtain the driving function. It is important to note that the NTIP process often produces sequences of bonds that include successive bonds that are not directly connected, resulting in discontinuities in the associated driving function. Fig.~\ref{fig:trees}a illustrates a NTIP tree, showing the order in which the bonds where invaded.

As mentioned, besides the order obtained from the NTIP model, there are many other ways to explore the same tree. In the following, we explain two additional exploring methods we have investigated.\\

\noindent\textit{Depth--First Search (DFS):} 
This algorithm explores each branch of the tree as deeply as possible before backtracking. In our approach, when reaching a splitting point while exploring the tree, the right--first strategy is used to decide the next branch to follow. 
Specifically, from the perspective of the direction from which you arrived at the splitting point, you should follow the branch that is furthest to the right. 
Here too, each time you backtrack from one branch to another,  a discontinuity appears in the driving function. Figure~\ref{fig:trees}b presents a NTIP tree, ordered according to the right--first DFS process.\\

\noindent\textit{Breadth--First Search (BFS)}:
This algorithm explores all sites at the same depth level, or chemical distance, before advancing the next depth level. The process begins at the root site, which is marked as visited. A queue is initialized by adding the sites of the tree that are neighbors of the root.
The algorithm then proceeds as follows: the site at the front of the queue is removed from the queue, marked as visited, and all unvisited neighbors of this site are added to the queue. This process continues systematically until all sites in the tree have been visited. 
Note that, when multiple branches exist at the same depth level, the algorithm jumps between branches. As a result, the driving function exhibits many discontinuities. 
Figure~\ref{fig:trees}c illustrates the visiting order obtained from the BFS process applied to the NTIP tree.

\subsection{The SLE theory \label{subsec:conformal_mapping}}

The SLE describes two--dimensional curves by parametrizing them with a time $t$, beginning at the origin $t=0$ and evolving up to infinity with $t\rightarrow\infty$. 
Consider a curve  $\gamma_t$ that grows in the upper half--plane $\mathbb{H}:\{ z = x+iy \in \mathbb{C}, y>0 \}$. 
The evolution of this curve during time $t$ is described by the conformal map  $g_t(z)$, which satisfies the Loewner equation~\cite{kennedy2007fast},
\begin{equation}
\partial_{t}g_{t}(z) = \frac{2}{g_{t}(z)-U_t},
\label{Eq:Loewner_equation}
\end{equation}
where $g_{0}(z)=z$. The term $U_t$ corresponds to a real valued function, known as driving function, which encapsulates all the topological properties of the curve. 

The driving function $U_t$ is obtained by finding a sequence of maps $g_t(z)$ such that each point of the curve $\gamma_t$ in $\mathbb{H}$ is mapped to the plane itself.
This map can be computed using the zipper algorithm~\cite{kennedy2009numerical}, where the driving function is approximated by a function that is constant over intervals of time $\delta t$. As a result, the driving function behaves like a discontinuous piecewise constant function.
Using a vertical slit discretization, the map is given by~\cite{kennedy2009numerical, ABRIL2024130066}:
\begin{equation}
g_{t}(z)=\sqrt{\left(z-U_t\right)^2+4\delta t}+U_t,
\label{eq:map_gt}
\end{equation}
where $\delta t_i = t_i - t_{i-1} = (Im\{z_i\})^2/4$ and $U_t = Re\{z_i\}$.  The functions $Im\{ \}$ and $Re\{ \}$ correspond to the imaginary and real parts, respectively. 

Each iteration in the algorithm is performed as follows: 
consider a curve $\gamma_t$ with $N$ points, with coordinates denoted by $z_i$. At time $t=0$, the driving function is set to $U_0=0$. The map $g_t$, which is the solution to the Loewner equation Eq.~(\ref{Eq:Loewner_equation}), is applied to each $z_i$.
After the first iteration, the new curve has $N-1$ points in the upper half--plane, as one point is already mapped to the real axis. In each subsequent iteration, the map $g_t$ is updated, and this procedure continues until the last point of the curve $z_i$ is mapped to the real axis, 
\begin{equation}
    z_i(t) = g_t(z_i(t - 1)) \circ g_{t-1}(z_i(t - 2)) \circ \cdots \circ g_{t_1}(z_i(0)). 
\end{equation}
For a process to be SLE, the driving function should obey $U_t = \sqrt{\kappa} B_t $, where $B_t$ is a one--dimensional Brownian motion with zero mean and variance $\kappa t$~\cite{rohde2011basic}. 
Conversely, from the driving function  $U_t$ one can get the curve $\gamma_t$  by applying the inverse conformal mapping,
\begin{equation}
\gamma_t= g_t^{-1}\left( U_t \right).
\label{Eq:gamma_inverse}
\end{equation}
In this way, a new trace is obtained with the inverse mapping.
The  discontinuities in the driving function cause the inverse process to generate a tree~\cite{cardy2005sle}.

\section{Results and Discussion\label{sec:results}}

On rectangular lattices in the upper half--plane with dimensions $1999\times 1000$ we simulated trees using the NTIP algorithm. 
As we show next, the branching structure of each tree can be mapped into several distinct driving functions, depending on the order in which the branches are explored. In addition to the order derived from the NTIP process itself, we employed the right--first Depth--First Search (DFS) and the Breadth--First Search (BFS) algorithms to generate alternative exploration orders. The driving function for each configuration was computed using the zipper algorithm with a vertical--slit map.

We divided the results into three parts:
The first part discusses the relationship between the driving functions and the Brownian motion, focusing on their statistical distributions and the computation of the power spectral density exponent.
The second part addresses the inverse problem, reconstructing the tree from the driving function. The new trace is compared to the original tree by analyzing their fractal dimensions. Finally, we focus on the DFS method to test if it is consistent with the SLE theory. 
\begin{figure*}[t]
\centering
\includegraphics[width=\linewidth]{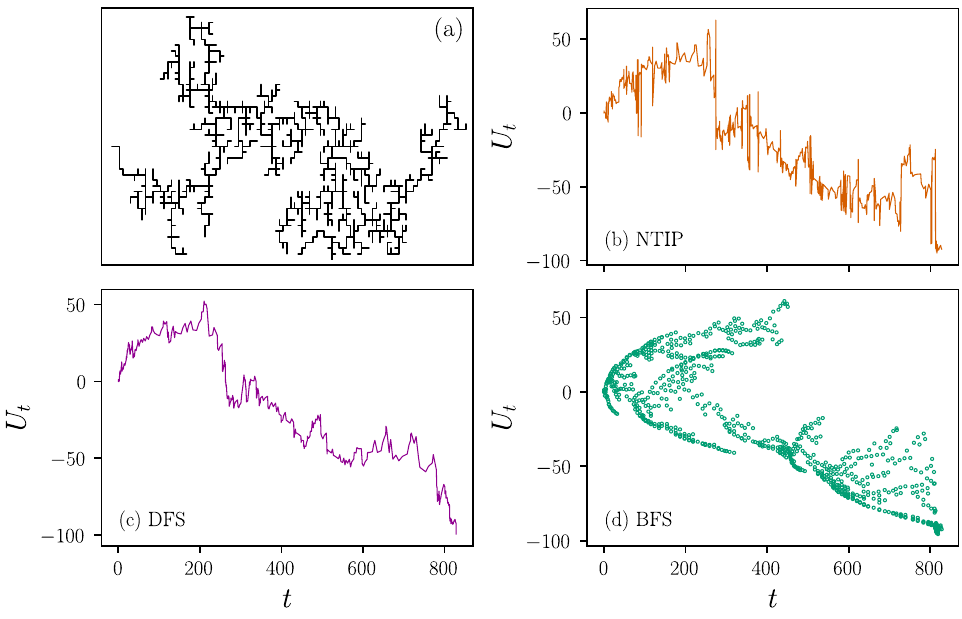}
\caption{Driving functions for three explorations of the same tree. The driving functions are computed using the zipper algorithm with a vertical slit. In (a), the studied tree contains $922$ sites. The driving functions are determined by the order in which the sites were visited: (b) according to the NTIP algorithm, (c) using  DFS with the right--first strategy, and (d) following the BFS method.}
    \label{fig:driv_functions}
\end{figure*}

\begin{figure*}[t]
\centering
\includegraphics[width=\linewidth]{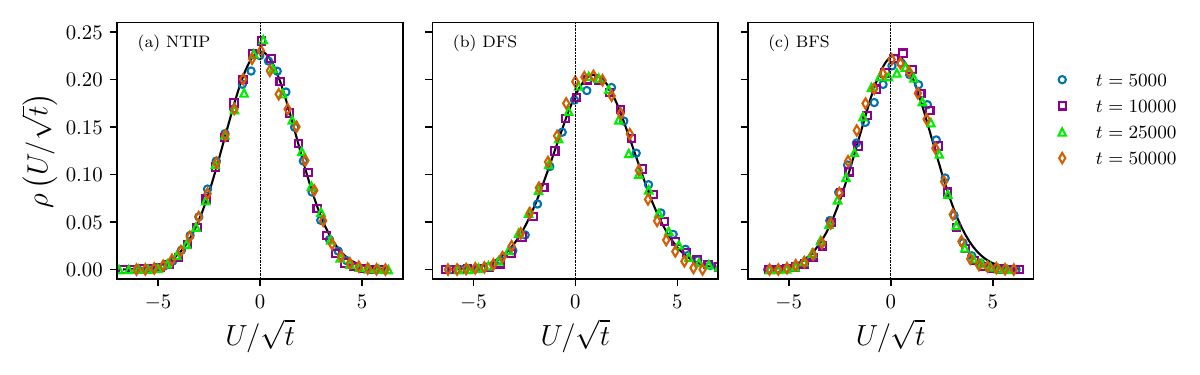}
\caption{ Normalized distributions of driving functions at four different times for each of the three methods. The driving functions $U_i$ were determined by assuming a piecewise constant behavior. Solid lines represent the Gaussian fits of the complete distribution for the four times. The coefficients of determination for the Gaussian fits are (a) $r^2=0.997$, (b) $r^2=0.999$ and (c) $r^2=0.991$.
We used over a total of 3000 samples for each method.
}
\label{fig:distr_tiempos}
\end{figure*}
\subsection{Statistics of driving functions}
We computed the driving functions $U_t$ for each method and investigated their properties.  For instance, we generated a tree with $922$ sites, as shown in Fig.~\ref{fig:driv_functions}a, and explored it using the three methods mentioned above. 

The driving function for the NTIP method, shown in Fig.~\ref{fig:driv_functions}b, exhibits multiple discontinuities, which occur when the tree grows along one branch and abruptly switches to another one. The driving function obtained from the DFS exploration to the tree is shown in Fig.~\ref{fig:driv_functions}c.  The right--first strategy introduces an asymmetry in the driving function. Discontinuities in the driving function appear during backtracking, causing the driving function to decrease. The driving function for BFS is shown in Fig.~\ref{fig:driv_functions}d. Here, we have a discontinuity after each iteration, as one jumps from one branch to another. To present a cleaner picture, in Fig.~\ref{fig:driv_functions}d we did not connect the points separated by a jump. 

\begin{table*}[b]
\centering
\caption{
The four moments of the normalized distributions  $U/\sqrt{t}$: mean $\mu$, variance $\sigma$, skewness $s$ and kurtosis $k$, are computed for the three methods at four different times, as shown in Fig.~\ref{fig:distr_tiempos}. The standard error of each moment is calculated averaging over 3000 samples. 
}
\label{tab:table1}
\begin{tabular}{@{}llllll@{}}
\toprule
 Model&$t$&$\mu$ &$\sigma$&$s$ & $k$\\
\midrule
NTIP&$t_1$&-0.008(31)&2.791(72)&-0.042(45)&-0.364(89)\\
&$t_2$&0.000(29)&2.649(68)&	-0.063(45)&-0.235(89)\\
&$t_3$&0.020(31)&2.800(72)&	-0.053(45)&-0.288(89)\\
&$t_4$&0.014(31)&2.868(74)&	-0.039(45)&-0.341(89)\\
& & & & &\\
DFS &$t_1$&1.103(35)&	3.734(97)&	0.149(45)&	-0.024(89)\\
&$t_2$&1.092(35)&	3.620(94)&	0.140(45)&	-0.130(89)\\
&$t_3$&0.920(35)&	3.754(97)&	0.143(45)&	-0.163(89)\\
&$t_4$&0.791(33)&	3.214(83)&	-0.058(45)&	-0.338(89)\\
& & & & &\\
BFS&$t_1$&0.180(31)&	2.821(72)&	-0.124(45)&	-0.478(89)\\
&$t_2$&0.213(30)&	2.665(68)&	-0.148(45)&	-0.404(89)\\
&$t_3$&0.141(31)&	2.810(72)&	-0.105(45)&	-0.435(89)\\
&$t_4$&0.096(31)&	2.722(74)&	-0.078(45)&	-0.396(89)\\
\botrule
\end{tabular}
\end{table*}

To compare with the properties of SLE, we investigate the relationship of the driving functions with Brownian motion. Plotting the distributions of different driving functions is a common method for analyzing stationarity and scaling properties. We computed 3000 driving functions with $L=1000$ for each method and studied their scaling behavior over times $t$.  
The normalized distributions $U_i/\sqrt{t_i}$ with $t_i=5000$, $10000$, $25000$, $50000$ for each method are shown in Fig.~\ref{fig:distr_tiempos}, and their first four moments in Table~\ref{tab:table1}. 
The mean, $\mu$, for the DFS and BFS methods is nonzero due to their preferential direction, by systematically going from right to left. In particular, the DFS process exhibits the largest drift in its mean. In contrast, the NTIP method does not display any directional bias. Meanwhile, the variance, $\sigma$, remains constant across the different times for each method. However, the DFS method displays a higher variance, indicating that it  diffuses more rapidly compared to the other methods. The skewness, $s$, suggests symmetry in the NTIP method, while, as previously noted, the other methods seem to contain a drift. Lastly, the kurtosis, $k$, indicates that DFS has tails that decrease more rapidly than those of a Gaussian distribution.  
For each method, a Gaussian fit is determined from the single distribution obtained by pooling the values at the four times. The coefficients of determination are all close to 1, indicating that the fit is appropriate. 

\begin{figure}[t]
\centering
\includegraphics[width=0.5\linewidth]{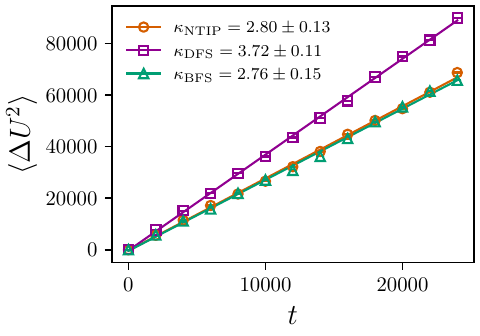}
\caption{Variance of driving functions over a fixed time for each method.
The slope of each curve corresponds to the diffusion coefficient $\kappa$. We averaged over a total of 3000 samples for each method, with error bars smaller than the symbols.}
\label{fig:variance_vs_t}
\end{figure}


We also analyzed the variance  of the driving functions, given by $\langle\Delta U^2\rangle= \langle U^2\rangle - \langle U\rangle^2$, 
at different times and found that it increases linearly with time, as shown in Fig.~\ref{fig:variance_vs_t}. We computed the slope, which corresponds to the diffusion coefficient $\kappa$, considering that the variance follows the relationship $\langle \Delta U^2\rangle=\kappa t$. 
Our results show that the coefficients for the BFS and NTIP methods exhibit similar values. As we noted, the mean values, $\langle U \rangle$, for these two methods are approximately zero, and thus, they do not significantly affect the variance. Conversely, for the DFS method, the  traversal approach  leads to faster diffusing function compared to the other methods, despite its higher mean, $\langle U \rangle$, being higher.

Another way to characterize the driving functions is through spectral techniques~\cite{shibasaki2020loewner}. In this approach, correlations are analyzed and quantified by the exponent of the power spectrum.\\

\noindent\textit{Power Spectral Density $S(f)$:} 
Considering that the driving function $U$  is constant in the intervals $t_i<t<t_{i+1}$ (in this section, we use $U=U_t$), its Fourier transform can be given by
\begin{equation}
\hat{U}(f) = - \frac{i}{2\pi f} \sum_{j=1}^{N} \left(U_{j}- U_{j-1} \right) e^{-2\pi i f t_j}.
\label{eq:}
\end{equation}
The power spectral density $S(f)$ is defined as the square of the magnitude of the Fourier transform of the driving function $U_t$,
\begin{equation}
S(f) = \frac{1}{N } \left| \hat{U} (f) \right|^2.
\end{equation}

The power spectral density sometimes follows the relationship $S(f)\sim f^{-\beta}$, where $\beta$ is an exponent that characterizes the scaling properties of the function. Typically, $\beta$ ranges from $0<\beta\leq 2$, where $\beta=2$ indicates that the driving function is an uncorrelated walk, which could correspond to a Brownian motion~\cite{shibasaki2020loewner, Krapf_2018}. Here, we employ the Lomb--Scargle periodogram, an appropriate method for computing the $S(f)$ of unevenly binned time series~\cite{VanderPlas_2018}.

\begin{figure*}[t]
\centering
\includegraphics[width=\linewidth]{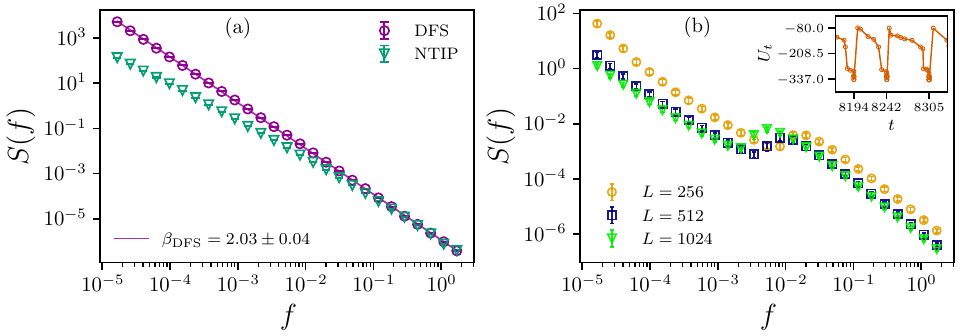}
\caption{Power spectral density $S(f)$ of the driving functions for the three methods. In (a), we show $S(f)$ of both the DFS and NTIP methods for a fixed lattice size with $L = 1000$.  $S(f)$ for the DFS method follows a power-law scaling across all frequencies, with $\beta\approx 2$, consistent with the scaling expected for Brownian motion. 
$S(f)$ for the NTIP method exhibits a curved profile in the double log scale, aligning with the DFS scaling at high frequencies but deviating from it in the low-frequency regime. In (b), $S(f)$ for the BFS method is shown using various lattice sizes $L$. The inset displays a sample of the driving function for $L = 512$. $S(f)$ was computed using the Lomb--Scargle method to account for uneven sampling in the time series. Error bars, computed by averaging over 3000 samples, are smaller than the symbols. }
\label{fig:psd_dfsNTIP}
\end{figure*}

Figure~\ref{fig:psd_dfsNTIP}a shows the results for DFS and NTIP. As depicted, only the DFS case exhibits a clear power-law relationship between $S(f)$ and frequency, with the exponent $\beta=2.03\pm 0.04$, obtained from the slope of the linear region in the log-log plot. In contrast, the $S(f)$ for NTIP case deviates from a power-law behavior.  
On the other hand, for the BFS case shown in Fig.~\ref{fig:psd_dfsNTIP}b, we do not observe any linear behavior. Since the shape of the spectrum differs significantly from the other cases, we compute $S(f)$ for different lattice sizes: $L = 256$, $512$, and $1024$. This case exhibits an intriguing behavior for $S(f)$ with a saddle--shaped spectrum, where the characteristic frequencies depend on the lattice size.
Examining the driving function obtained using the BFS algorithm (inset of Fig.~\ref{fig:psd_dfsNTIP}b) one observes, that at certain times, the function displays  larger discontinuities, increasing $U_t$, followed by a series of smaller discontinuities decreasing $U_t$. 
Each of these cycles corresponds to a shell of nodes of equal depth in the BFS. This cyclic behavior of $U_t$ is reflected in $S(f)$.

Considering the results for the distributions and the exponent $\beta$, it is possible that the driving function of DFS behaves like a Brownian motion with a slight drift. This finding motivates us to compute the diffusion coefficient using three additional numerical tests to verify if the obtained values are equal or not  as discussed next.

\subsection{Inverse Mapping: Trees from the driving function}

We used the three methods to generate driving functions and now our aim is to reconstruct the original tree from these driving functions. 

\begin{figure}[t]
\centering
\includegraphics[width=0.5\linewidth]{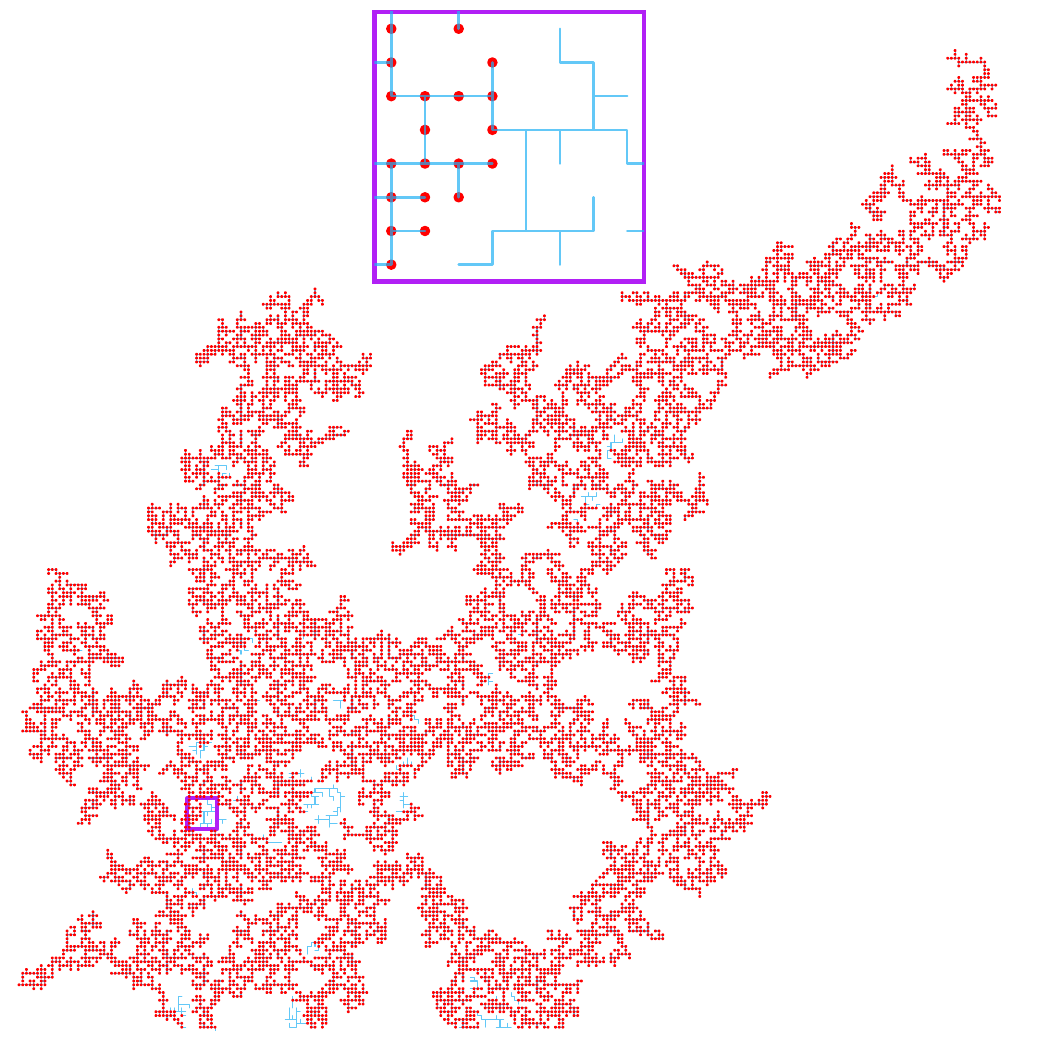}
\caption{Trace obtained from the driving function generated from a tree using the BFS method. The blue line represents the original generated tree, while the red sites indicate the reconstructed trace.  The zipper algorithm fails to recover all the sites; however, in this case, the reconstructed trace differs by only 
$1\%$ from the original.
The inset provides a zoomed view of the tree highlighting its details. The original cluster contains 25488 sites.}
\label{fig:new-trace}
\end{figure}

\begin{figure*}[b]
\centering
\includegraphics[width=0.5\linewidth]{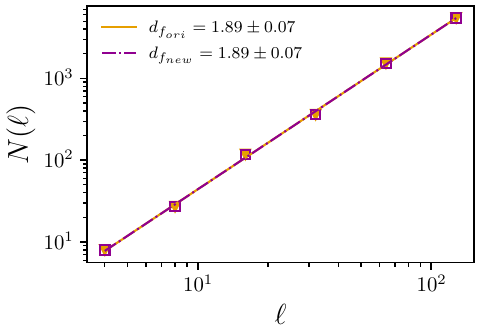}
\caption{Fractal dimension of the trees. The slope of the solid line represents the fractal dimension of a tree generated using invasion percolation. The slope of the dashed line is the fractal dimension of the new trees computed from the driving function. The tree contains $25488$ sites. }
\label{fig:df_trees}
\end{figure*}

Although the zipper algorithm is an efficient numerical tool for solving Loewner's equation, it has certain limitations. In direct mapping, it is known that, when the trace touches itself, the regions enclosed by this trace are mapped to the real axis~\cite{cardy2005sle}. 
Our branching structures are trees, and thus enclosed regions should not form; however, when the trace approaches itself, it can create nearly trapped regions. 
Formally, all points on the trace should remain above the real line until they are mapped by the Loewner evolution. Nonetheless, the imaginary part of these trapped points can become exceedingly small. 
Due to precision limitations in the numerical method, such points may be erroneously treated as if they were already on the real line, causing the time increment in the driving function to be recorded as zero. Consequently, during the inverse mapping, these misclassified points cannot be distinguished from the rest of the trace and fail to return to their original positions.

In Fig.~\ref{fig:new-trace}, 
we show an example of a tree from which we obtained a driving function using the BFS method. Using this driving function, we applied the inverse zipper algorithm, as described in Eq.~(\ref{Eq:gamma_inverse}) to reconstruct the sites of the original tree. 
The reconstructed sites are plotted in red, while the original tree is shown in blue.
As observed, the original tree is nearly fully recovered. However, due to the algorithmic limitations already described, not every point of the original tree is reconstructed. 
In fact, when using the BFS  method, approximately $1\%$ of the sites could not be recovered. 
Methods employing extended computational precision might help reduce the fraction of unrecovered sites. However, our tests using the \texttt{mpfr} library, which allows for 400 decimal places of precision, still revealed discrepancies at sufficiently large scales. 

The difference between the original tree and the one reconstructed from the driving function might be assessed by comparing their fractal dimensions. The fractal dimension, $d_f$, of the original and reconstructed trees can be computed using the sandbox method~\cite{TEL1989155}. 
Starting from the central point of the tree, boxes of size $\ell$ are created around the center, and the number of sites, $N(\ell)$, within each box is counted. This follows the scaling relationship $N(\ell) \sim \ell^{d_f}$. The curves of $N(\ell)$ and the corresponding $d_f$ values are shown in Fig.~\ref{fig:df_trees}. 
The fractal dimension of both the original and reconstructed trees, using the BFS method, are $d_{f_{ori}}=1.89\pm 0.07$. 
The reconstruction using the NTIP method produces a similar result, yielding the same fractal dimension for the reconstructed tree.
In contrast, the DFS method results in a fractal dimension of $d_f=1.83\pm 0.09$ for the reconstructed tree, reflecting a $3\%$ deviation from the fractal dimension for the original tree. This discrepancy arises due to a larger trapped region in the DFS method, which affects more sites compared to the other methods. In this way, we see that, despite the limitations of the zipper algorithm, both the BFS and NTIP methods effectively reconstruct the original tree with reasonably high precision.
\subsection{Numerical tests of SLE theory}
Here we will focus exclusively on the DFS method, since its statistical properties closely resemble those of a Brownian motion. Moreover, it can be viewed as a loopless path that begins at the origin and ends at the upper boundary.
We computed the diffusion coefficient from driving functions obtaining $\kappa_{dSLE}=3.72 \pm 0.11$, from Fig.~\ref{fig:variance_vs_t}. For SLE curves, the diffusion parameter is related to the fractal dimension via~\cite{beffara2008},
\begin{equation}
d_f=1+\frac{\kappa}{8},
\end{equation}
where the $d_f$ is the fractal dimension of the trace. The curve obtained from the right--first DFS has the fractal dimension of two--dimensional clusters, $d_f=1.89\pm0.07$. If it were an SLE trace, this fractal dimension would correspond to $\kappa_{d_f}=7.12\pm 0.56$.\\

\begin{figure*}[t]
\centering
\includegraphics[width=0.5\linewidth]{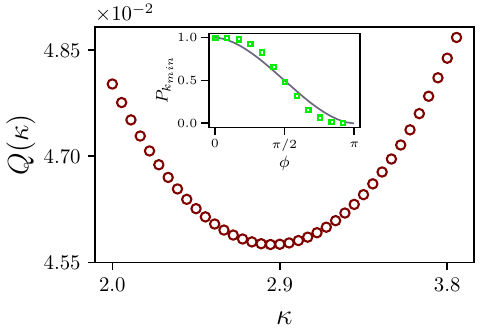}
\caption{Mean square deviation $Q_{\kappa}$ as a function of the parameter $\kappa$ for trees of lateral size $L_y=256$. We have analyzed $578$ trees, distributing points within the ranges $0<R<256$ and $0<\phi<\pi$, with steps of $10$ and $\pi/180$, respectively. The inset shows the probabilities evaluated at $\kappa_{min}$ for Schramm's probability $P_{\kappa}(\phi)$ represented by the solid line, and the probability obtained numerically $P(\phi,R=210)$ with the square markers. }
\label{fig:lpp}
\end{figure*}

\noindent\textit{Left Passage Probability (LPP):} It measures the probability that a curve passes to the left of a point $z=Re^{i\phi}$ in $\mathbb{H}$. This probability depends only on the angle $\phi$ and the parameter $\kappa$, and is given by Schramm's formula~\cite{schrammpercolation2001},
\begin{equation}
P_{\kappa}(\phi) = \frac{1}{2}+\frac{\Gamma\left(\frac{4}{\kappa}\right)}{\sqrt{\pi}\Gamma\left(\frac{8-\kappa}{2\kappa}\right)} \cot(\phi)\,\, _{2}F_{1} \left( \frac{1}{2}, \frac{4}{\kappa}, \frac{3}{2}, -\cot^2(\phi) \right),
\end{equation}
where $\Gamma$ is the Gamma function and $_2F_1$ represents the Gauss hypergeometric function. The diffusion coefficient $\kappa$ can be determined by comparing Schramm's probability $P_\kappa$ with the numerically computed probability $P(\phi,R)$, which is evaluated for various radii $R$ and angles $\phi$. The optimal value of $\kappa_{LPP}$ is obtained by minimizing the mean square deviation $Q(\kappa)$, defined as~\cite{daryaei2012watersheds},

\begin{equation}
Q(\kappa) = \frac{1}{N}\sum_{R} \sum_{\phi} \left[P(\phi,R)-P_{\kappa}(\phi)\right]^2,
\end{equation}

\noindent where $N$ is the total number of points $z$. We estimated the  minimum value of $Q(\kappa)$, as shown in Fig.~\ref{fig:lpp}, obtaining $\kappa_{LPP}=2.85\pm0.05$. \\

\noindent\textit{Winding angle}: This method examines the local angles between segments from the origin to the upper boundary, starting with $\theta_{0}=0$, and iteratively computes $\theta_{i+1}=\theta_{i}+\alpha_{i}$, where $\alpha_i$ is the angle between the line connecting points $i$ and $i+1$ and  the tangent line of the curve at point $i$. 
The variance of the winding angle follows the relationship~\cite{schrammpercolation2001},
\begin{equation}
\langle\theta^2\rangle = \frac{\kappa}{4}\ln(L_y) + b,
\label{Eq:windingangle}
\end{equation}
where $L_y$ is the vertical axis and $b$ is a constant.

\begin{figure}[b]
\centering
\includegraphics[width=0.5\linewidth]{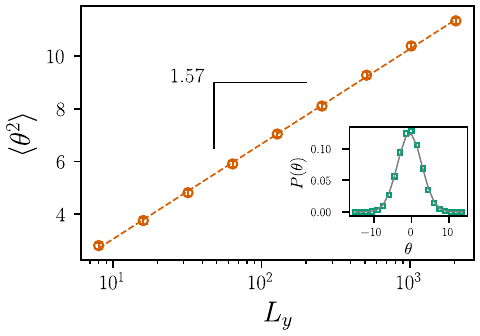}
\caption{Variance of winding angles as a function of the lateral size $L_y$ for paths generated using the DFS method on a tree, including backtracking sites. A total of 1000 trees of size 4096 were generated, until they reached the upper boundary. The inset shows the distribution of the angles for trees with  $L_y=2048$. The slope $\kappa_{\theta}/4$ corresponds to the diffusion coefficient, yielding $\kappa_{\theta}=6.29\pm0.05$.}
\label{fig:winding_angle}
\end{figure}

Instead of using the DFS method, where each site is recorded only once, here we modified our approach to also include the backtracking sites thus avoiding jumps. Using this method, we computed the variance and distribution of the winding angles. The dependence of the variance on the lateral size $L_y$ is shown in Fig.~\ref{fig:winding_angle}, along
with the angle distribution for $L_y=2048$.

As discussed, the driving function for the DFS method seem to be a Brownian motion with a slight drift, characterized by a nonzero mean. According to Schramm's criteria~\cite{schrammpercolation2001,werner2003randomplanarcurvesschrammloewner}, such paths lacks conformal invariance due to the preferential direction introduced by the drift. Additionally, we compared the diffusion coefficients obtained using the DFS method with four different methods, as summarized in Table~\ref{tab:table2}. The lack of conformal invariance, combined with the fact that these diffusion coefficients are not equal, demonstrates the inconsistency of DFS trees with SLE. Similar deviations have also been observed in other systems, as reported in \cite{pose2018schramm} and \cite{ABRIL2024130066}.

\begin{table*}[h]
\centering
\caption{
Diffusion coefficients for the DFS method from invasion percolation, calculated using various numerical tests of the SLE theory.} \label{tab:table2}%
\begin{tabular}{@{}cccc@{}}
\toprule
$\kappa_{d_f}$ & $\kappa_{\theta}$ &$\kappa_{LPP}$&$\kappa_{dSLE}$ 
\\
\midrule
$7.12\pm 0.56$ & $6.29\pm0.05$&$2.85\pm 0.05$ & $3.72\pm 0.11$\\
\botrule
\end{tabular}  
\end{table*}

\section{Conclusions\label{sec:conclusions}}
We simulated tree-like structures and explored them using three methods: NTIP, DFS, and BFS. The statistical properties of their driving functions indicate that their distributions share some characteristics with the Brownian motion. Notably, the spectral density analysis showed that only the driving function generated by the DFS method exhibits a spectral exponent, suggesting behavior similar to a Brownian motion.

Furthermore, we investigated the inverse problem, where new traces were obtained from the driving functions of the three methods. The trace most similar to the original tree was obtained using either the BFS or the NTIP method, as evidenced by its fractal dimension. However, we encountered some numerical limitations of the zipper algorithm, particularly regarding the precision of the computed driving function.

Finally, we examined paths excluding their jumps to assess their convergence with the SLE theory. In particular, we focused on loopless paths generated by the DFS method, since the BFS and NTIP methods are incompatible with SLE, given that their driving functions do not seem to be Brownian motions.
We found that the diffusion coefficients of these paths do not converge and, therefore, can not be described by the SLE theory.

\bmhead{Acknowledgments}
We thank the Brazilian agencies CNPq, CAPES, FUNCAP, the National Institute of Science and Technology for Complex Systems (INCT-SC) in Brazil for financial support.

\bmhead{Data availability}
The data that support the findings of this study are available within the article.
\bibliography{sn-bibliography}

\end{document}